\documentclass[12pt,onecolumn]{article}
\usepackage{times}
\usepackage{graphicx}
\usepackage{url}
\usepackage[pagebackref,breaklinks=true,colorlinks,bookmarks]{hyperref}
\hypersetup{
     colorlinks   = true,
     citecolor    = blue
}
\usepackage{verbatim} 
\usepackage{amsthm, amssymb, amsfonts}
\usepackage{mathtools}
\usepackage[american]{babel}
\usepackage{fancyhdr}
\usepackage{parskip}
\usepackage[utf8]{inputenc}
\usepackage[T1]{fontenc}
\usepackage{lastpage}
\usepackage[letterpaper, left=1in, right=1in, bottom=1in, top=1in]{geometry}
\usepackage{tabulary}
\usepackage[font={small,it}]{caption}
\usepackage{setspace}
\usepackage{url}
\usepackage{sectsty}
\subsubsectionfont{\fontsize{14}{15}\selectfont}
\subsectionfont{\fontsize{15}{15}\selectfont}
\begin{document}

\title{\textbf{2001-2013: Survey and Analysis of\\ Major Cyberattacks}}

\author{
Tavish Vaidya\\
Georgetown University\\
\date{}
}

\maketitle

\singlespacing
\begin{abstract}

Widespread and extensive use of computers and their interconnections in almost all sectors like communications, finance, transportation, military, governance, education, energy etc., they have become attractive targets for adversaries to spy, disrupt or steal information by presses of keystrokes from any part of the world. This paper presents a survey of major cyberattacks from 2001 to 2013 and analyzes these attacks to understand the motivation, targets and technique(s) employed by the attackers. Observed trends in cyberattacks have also been discussed in the paper.

\end{abstract}

\section{Introduction} 
\label{sec:introduction}

Cyberattacks are computer-to-computer attacks undermining the confidentiality, integrity, and/or availability of computers and/or the information they hold\cite{cyberattack_definition}. Computer networks have no geographical borders that need to be crossed for an attacker to steal information. This grants freedom to any attacker to pick his target anywhere in the world and carry out a cyberattack. Therefore, securing computer systems is as important as securing physical entities from being attacked. In terms of money, Ponemon Institute\cite{cost_study} estimated the average cost of cyberattacks to be \$11.6 million per organization for 2013, which was 26 percent more than 2012.

Cyberattacks have not only caused losses in billions of dollars\cite{total_global_cost_report}, but also had psychological impact on human psyche. As an example, in August 2004, fear of cyberattacks during Olympic games in Greece kept people from attending the Olympic events\cite{fear_cyberattack}. With Internet of Things\footnote{http://en.wikipedia.org/wiki/Internet\_of\_Things} already here, securing computer networks and end devices becomes a paramount concern to prevent cyberattacks from disrupting and hijacking them for malicious purposes. Hence, it imperative and necessary to understand the motivations, attack vectors and weaknesses exploited by past cyberattacks. Lessons must be learned from past experiences to improve upon all aspects necessary for defending against cyberattacks in future.

This paper covers major cyberattacks starting from 2001 till 2013. The scope of this paper is limited to cyberattacks that caused significant monetary losses, threatened critical infrastructure or national security, had potential to cause loss of life and damage to physical property or involved data-leaks exposing personal information of users. This paper also covers cyber-espionage campaigns and cyberattacks with political motivations or agenda as well as some acts of hacktivism. It then examines the trend in attack methodology, frequency, motivation behind the attacks, damage caused, attribution and if any lessons were learned from past cyberattacks to improve security of computer systems. 

\textit{(The source of the cyberattacks have been mentioned when the attackers were identified or suspected. In attacks where the source was unknown, we omit mentioning this fact explicitly throughout the text of this paper. The term attack(s) will be used interchangeably with cyberattack(s), unless explicitly mentioned otherwise.)}

\section{Revisiting past Cyberattacks} 
\label{sec:revisitng_past_cyberattacks}

This sections surveys past cyberattacks to put the facts together and help support the analysis and uncover trends. Looking at each attack in isolation only provides limited information, however, cyberattacks can also be related to one another and therefore, can provide a lot more understanding when analyzed together.

This paper classifies cyberattacks based on their targeting, i.e. undirected or directed/targeted. Undirected cyberattacks are not directed towards a specific target but attack any vulnerable host. Directed/targeted attacks are carried out against specific targets and are designed to exploit specific weaknesses of the targeted systems.

\subsection{Undirected Cyberattacks} 
\label{subsec:undirected_cyberattacks}
The first undirected attack within the scope of this paper was the Anna Kournikova virus identified in February, 2001\cite{2001_anna_news}. It exploited multiple vulnerabilities in Windows operating system and Microsoft Outlook to spread to other systems\cite{2001_anna_casestudy}. Dutch programmer Jan de Wit created the virus to see if lessons were learned from the {ILOVEYOU}\cite{iloveyou_wiki} virus from last year\cite{2001_anna_motivation}. 
In July 2001, self-propagating \textit{Code Red} worm\cite{2001_code_red_news} infected 359,000 computers in less than 14 hours by exploiting a buffer overflow vulnerability in Microsoft IIS Server, disrupting hosted websites\cite{2001_code_red_damage}. The estimated losses were put close to \$2.6 billion\cite{2001_code_red_cost}.

In January 2003, \textit{Slammer} worm wrecked havoc on the Internet by flooding networks with queries causing the routers to collapse\cite{slammer_worm_news}. \textit{Slammer} worm also exploited a buffer overflow bug in Microsoft SQL Server\cite{slammer_worm_analysis}, for which a patch was already available 6 months before the worm was launched. 
Same year in August, \textit{Blaster} worm infected more than 48,000 computers worldwide and caused a distributed denial-of-service(DDoS) attack on windowsupdate.com\cite{blaster_worm_news}. Author of the \textit{Blaster} worm, Jeffrey Lee Parson\cite{blaster_worm_attribution}, exploited vulnerabilities in the Microsoft Remote Procedure Call Interface to infect vulnerable hosts\cite{blaster_worm_cert} even though patches were released a month before the attack. 
Also in August 2003, \textit{Sober} email worm\cite{sober_worm_news} was used to send political spam\cite{sober_worm_hate} and in 2005, its variants were circulated with fake emails impersonating the FBI and the CIA\cite{sober_worm_fbi_cia}. 

In 2004, \textit{Mydoom} worm\cite{mydoom_worm_news} caused an estimated loss of \$38.5 billion\cite{mydoom_worm_damage_loss}. Believed to have originated in Russia\cite{mydoom_worm_attribution}, the worm infected more than 500,000 machines and was sent out as email attachments\cite{mydoom_worm_damage_loss}, with later version exploiting a zero-day vulnerability in Internet Explorer browser\cite{mydoom_worm_zero_day}. 
Sasser worm\cite{sasser_worm_news} disrupted services of many companies in 2004\cite{sasser_worm_attribution}, that claimed losses totaling \$155,000 in civil lawsuits\cite{sasser_worm_losses}. \textit{Sasser} exploited a known buffer overflow vulnerability in Microsoft's Local Security Authority Subsystem Service\cite{sasser_worm_exploit} and was attributed to Sven Jaschan\cite{sasser_worm_attribution}. 
Just before the Christmas holidays of 2004, \textit{Santy} worm was seen using search engines to find servers running vulnerable phpBB software and was able to deface over 40,000 websites\cite{santy_worm_news,santy_worm_damage}.

In August 2005, with monetary benefit being the primary motive, Farid Essebar and Atilla Ekici\cite{Zotob_worm_motive} launched the \textit{Zotob} worm\cite{zotob_worm_news} that exploited known vulnerabilities in Windows 2000 operating system\cite{Zotob_worm_exploit}. 
\textit{Zotob} slowed down computers of more than 100 companies causing them to continually crash and reboot while also opening a backdoor\cite{Zotob_worm_effect}. The worm caused an estimated average loss of \$97,000 and 80 hours of cleanup per affected company\cite{Zotob_worm_losses}.

In November 2008, \textit{Conficker} worm was detected exploiting a vulnerability present in multiple Microsoft operating systems that allowed arbitrary remote code execution, for which Microsoft had issued a critical security bulletin on 23\textsuperscript{rd} October, 2008\cite{conficker_worm}. \textit{Conficker} infected 11 million hosts globally\cite{conficker_infected_numbers} with an estimated economic cost of \$9.1 billion\cite{conficker_reported_cost}. \textit{Conficker} is believed to have originated in Ukraine\cite{conficker_attribution}.

Undirected cyberattacks, exploiting vulnerabilities in widely deployed software, remain to be major threats. \textit{Heart Bleed} vulnerability is a recent witness\cite{heartbleed_1,heartbleed_2}.

\subsection{Targeted/Directed Cyberattacks} 
\label{subsec:targeted_directed_cyberattacks}

Targeted cyberattacks have been further categorized according to their targets and potential \mbox{motivations}.

\subsubsection{Cyberattacks directed towards Nations} 
\label{subsec:cyberattacks_on_nations}

Cyberattacks have targeted nations by specifically going after targets within a particular nation and disrupting normal operations of computers and networks.

The US-China spy plane incident in April, 2001 led to a month long online battle between US and Chinese hackers, both causing defacements and posting messages on government related websites while accusing each other of the incident\cite{2001_china_US_hacking}.

April 2007 witnessed the first series of cyberattacks targeting a particular nation, Estonia\cite{estonia_news}. Botnets from all around the world were directed towards Estonia in a Distributed Denial-of-Service(DDoS) attack and also posted messages on various defaced websites. 
The main targets were the websites of Estonian President and parliament, government ministries, political parties, news organizations, the two biggest banks and telecommunication firms\cite{estonia_news}. Consequently, Estonia had to cut off its networks from the outside internet to protect against these attacks. Only one bank reported an estimated loss of 1 million dollars\cite{estonia_cost_impact}. The attacks were seen as patriotic response and blunt payback from Russia against the Estonian government's decision to relocate the statue of Bronze Soldier of Tallinn, an elaborate Soviet-era grave marker, as well as war graves in Tallinn. Some reports alleged Russian government involvement, given the money and technical skills required to carry out such a sophisticated and co-ordinated attack on a country\cite{estonia_news2,estonia_russia_accused}. Other experts doubted Russian involvement\cite{estonia_news_attribution_noevidence}\cite{estonia_attribution_not_russia}. In 2009, a Kremlin-backed youth group claimed to have carried out these cyberattacks\cite{estonia_real_claim}.

In September 2007, Israel disrupted Syrian air defense systems during Operation Orchard allowing Israeli F-15s and F-16s to enter Syrian airspace without detection\cite{2007_israel_syrian_hack_news,2007_israel_syrian_hack}. 

August 2008, Georgia suffered massive DDoS attacks and traffic re-routing that crippled its limited Internet infrastructure\cite{2008_aug_georgia,2008_aug_georgia_out}. The attacks started before the beginning of conventional war between Russia and Georgia and continued alongside the military engagement. Georgia blamed Russia for the cyberattacks, though the attacks originated from infected computers in various countries\cite{2008_aug_georgia_motive}.

In January 2009, Israeli websites belonging to small companies and government bodies including the Israeli Defense Forces and the Israel Discount Bank were targeted with DDoS attacks and defacements in protest and retaliation to Israeli military attacks on Gaza\cite{2009_israel_news2}. Israel suspected former Soviet Union hackers, paid by Hamas or Hezbollah, for carrying out the attacks\cite{2009_israel_news}. 
In July 2009, DDoS attacks were directed at major government, news media, and financial websites in South Korea and the US\cite{2009_july_ddos_US_SK,2009_july_ddos_US_SK2}. South Korea, where some websites suffered outages for days, blamed the North Korean telecommunications ministry for the attacks\cite{2009_july_ddos_US_SK_attribution}.

In October 2012, DDoS attacks on Iran slowed down the Internet throughout the country\cite{2012_oct_iran_ddos}.

\subsubsection{Cyberattacks threatening National Security} 
\label{subsec:cyberattacks_threatening_national_security}

Government organizations and personnel, military networks, defense contractors and other entities tied to national security have been targets of various cyberattacks worldwide aimed at getting sensitive information on military, political, economic, strategic and government.

Beginning in 2003, several US government agencies, including the departments of State, Energy and Homeland Security, NASA, as well as defense contractors were targeted by a series of coordinated cyberattacks\cite{titan_rain_news1,titan_rain_times_article}. The attacks, code named \textit{Titan Rain}, breached hundreds of unclassified networks siphoning off any available information. In August 2005, SANS Institute revealed that the attacks originated in Chinese province of Guangdong\cite{titan_rain_attribution}.

In May 2006, hackers targeted US State Department's headquarters and offices dealing with Asia, breaching the unclassified network\cite{state_dept_news}. The attacks exploited a zero-day vulnerability in Microsoft operating system and the malware exploit was delivered via phishing emails\cite{state_dept_exploit}. Sensitive information including passwords were believed to have been stolen.\cite{state_dept_news,state_dept_damages}.
In August 2006, Maj. Gen. William Lord publicly stated that 10 to 20 terabytes of data has been downloaded by China from NIPRNet\cite{2006_red_storm_rising}. 
Also in 2006, spyware was found on computer systems of classified departments at China's China Aerospace Science \& Industry Corporation\cite{china_attacked}.

In June 2007, cyberattacks originating in China targeted the US Department of Defense\cite{dod_news_2007_attribution}. Spoofed email with recognizable names and malicious code was sent to the office of the Secretary of Defense. The malicious code exploited a known vulnerability in Microsoft Windows operating system. Sensitive information accessible from the network, like user IDs and passwords that allowed access to the entire unclassified network\cite{dod_news_motive,dod_news}, was exfiltrated, causing 1,500 computers to be taken offline\cite{dod_news_damage_nbc}. 
In October, China accused foreign hackers from Taiwan and the US for stealing information, without providing any other details\cite{china_attacked}. 
In November 2007, about 1,100 employees of nuclear arms lab at Oak Ridge National Laboratory were targeted with phishing emails with attached malware, originating at Internet and web addresses located in China. The attackers were able to obtain visitor information to the lab since 1990\cite{2007_nuclear_lab_news}.

During the summer of 2008, the databases of Republican and Democratic presidential campaigns containing sensitive internal documents and private data were copied in a cyberattack\cite{2008_campaign_hacked}. The attacks were traced back to China by US intelligence agencies\cite{2008_campaign_attribution}. 
In November 2008, classified and unclassified networks of US Department of Defense and US Central Command were hacked because of a SillyFDC malware variant which was delivered via an infected USB stick at a base in Middle-East\cite{2008_central_command_real,2008_central_command}.

In 2009, \textit{Conficker} worm infection grounded French fighter planes\cite{conficker_impach_frenchplanes} and computers on board Royal Navy warships and submarines were also affected\cite{conficker_navy_news}. 
In March 2009, a global cyber-espionage network named \textit{GhostNet} was revealed, which exploited a known vulnerability in Adobe PDF reader\cite{2009_ghostnet_exploit}. 
\textit{GhostNet} spied on multiple high-value targets like ministries of foreign affairs, embassies etc. in 103 countries, international organizations, news media and NGOs\cite{2009_ghostnet_news2}. Most attacks under \textit{GhostNet} originated in China, though involvement of the Chinese government was not ascertained\cite{2009_ghostnet_news}.
In April 2009, hackers stole terabytes of data related to design and electronics systems of the F-35 Lightning II fighter jet. The sensitive data was encrypted before exfiltration to sources in China, making it impossible to determine precisely what information was stolen\cite{2009_fighter_jet}.

In April 2010, computer systems of the Indian Defense Ministry and Indian embassies in various countries were compromised\cite{2010_indian_defense_news}. Attacks stole classified information including designs of weapon systems, internal security assessments of sensitive regions and emails from Dalai Lama's office\cite{2010_indian_defense_damage}. The attacks were traced back to China\cite{2010_indian_defense_attribution,2010_indian_defense_news}.

In April 2011, within a month of the RSA breach\cite{2011_rsa_news}, the stolen SecurID tokens were used to hack defense contractor L-3 Communications for theft of sensitive information\cite{2011_l3_news}. In May, Lockheed Martin\cite{2011_lockheed_news,2011_lockheed_exploit} as well as Northrop Grumman\cite{2011_northrop_news,2011_rsa_news} were targeted in a cyberattack using the stolen RSA SecurID tokens, though the attacks were thwarted. 
In July 2011, unknown attackers breached Pentagon networks stealing 24,000 files, with the exact damage being undisclosed\cite{2011_pentagon_news,2011_pentagon_motivation}. 
In August 2011, operation \textit{Shady Rat} was revealed to have been attacking 70 corporations and government organizations in the US since mid-2006 and other international targets\cite{2011_shady_rat_news,2011_shady_rat_report}. This cyber-espionage campaign employed spear-phishing with attached files containing malware that exploited a known vulnerability in Microsoft Excel to open a backdoor\cite{2011_shady_rat_symt}. 
In October 2011, 2 US satellites were interfered with for few minutes, allegedly by attackers from China\cite{2011_satellite_hack}. 

In September 2012, the White House became a victim of spear-phishing attacks allegedly carried out by hackers in China\cite{2012_white_house}. In December 2012, computer networks of Indian government were breached, compromising 10,000 email accounts of top government officials and information on troop deployment\cite{2012_dec_indiahack}. 

In February 2013, a US government report revealed that 23 US gas companies were targeted by cyberattacks that stole potentially security-sensitive information\cite{2013_feb_gas_report}. 
In May, infiltration of systems at defense contractor QinetiQ by Chinese hackers was discovered. The attackers were in the system since 2007 because of a known security flaw resulting in the compromise and exfiltration of most of the company's research\cite{2013_may_qinetiq}.
Also in May, a report by the Defense Science Board reported that designs of US defense systems including the Patriot missile system (PAC-3), Terminal High Altitude Area Defense and Navy’s Aegis ballistic-missile defense system had been compromised by persistent, highly-sophisticated cyberattacks carried out by China\cite{2013_may_us_china_spying}. 
In August 2013, hackers gained access to personal information, social security numbers and payroll information of 14,000 current and former employees at the US Department of Energy\cite{2013_aug_doe}. 
In September 2013, \textit{Operation Kimsuky} was revealed to be spying and stealing information from South Korean think-tank organizations using malware, delivered via a spear-phishing campaign. North Korea was blamed for the targeted attack as the malware specifically disabled a particular South Korean antivirus\cite{2013_sep_kimsuky}.

\subsubsection{Cyberattacks on Companies and Organizations} 
\label{subsec:cyberattacks_on_companies_and_organizations}

Cyberattacks discussed in this section targeted organizations in banking, finance, oil and energy, communications, technology, news media, retail sectors and other private enterprises.

In January 2001, Microsoft websites were targeted with a DDoS attack for a day due to poor configuration of the network\cite{2001_microsoft_ddos}. 

In 2004, hosts infected with the \textit{Mydoom} worm were used in DDoS attacks on Google, Microsoft and other companies\cite{mydoom_worm_damage_google}.

In January 2007, retail giant TJX suffered a targeted hack due to poor security of their {WiFi} network that allowed sniffing of information\cite{2007_jan_tjx_exploit} on 45.7 million accounts including credit and debit card numbers\cite{2007_jan_tjx_news,2007_jan_tjx_numbers}. Albert Gonzalez was convicted and sentenced to 20 years in prison for being the primary hacker\cite{2007_jan_tjx_attr} in the attack that costed \$4.5 billion\cite{2007_jan_tjx_cost}.	

During the summer of 2008, 3 US oil companies were targeted with phishing emails containing links to spyware to infect their systems. Attackers stole data on discoveries of new oil deposits, e-mails, passwords and other information related to executives who had access to proprietary exploration and discovery information\cite{2008_oil_companies_attribution,2008_oil_companies}. China's involvement was suspected in the attacks\cite{2008_oil_companies_attribution}.

In December 2009, Citibank lost tens of millions of dollars to a cyberattack, which it denied\cite{2009_citibank_news1,2009_citibank_news2}. Hackers used spyware keylogger in one of the publicly known incidents to gain access to user account\cite{2009_citibank_userhack}.
Source IPs used in the attacks had been linked to Russian Business Network hacking group in the past\cite{2009_citibank_attribution}.



In January 2010, Google announced that it has been targeted by sophisticated cyberattacks that compromised various user accounts\cite{2010_opaurora_google_blog}. At least 34 other companies including Yahoo, Symantec, Adobe, Northrop Grumman and Dow Chemical were hit by similar attacks, named \textit{Operation Aurora}. The attacks exploited a zero day vulnerability and other known vulnerabilities in Microsoft Internet Explorer and used spear-phishing to deliver malware for stealing data from targeted companies\cite{2010_opaurora_exploit_damage,2010_opaurora_exploit2,aurora_report}. Earlier reports \cite{2010_opaurora_exploit_damage,2010_opaurora_exploit2} mentioned that a zero-day vulnerability in Adobe Reader might have been exploited, but no conclusive evidence was found\cite{no_evidence}.
In October 2010, it was reported that banks in the US lost over \$12 million to hackers who used Zeus trojan to infect computers via phishing emails and recorded keystrokes to steal bank account credentials\cite{2010_global_bank_hacks_exploit}. 100 people were charged as suspects.\cite{2010_global_bank_hacks}. 

In March 2011, Epsilon suffered a data breach due malware delivered via spear-phishing campaign\cite{2011_epsilon_news,2011_epsilon_motivation,2011_epsilon_exploit}. The breach cost \$225 million in damages plus additional costs to clients with the total running in billions\cite{2011_epsilon_costs}. 
Also in March, security firm RSA suffered a massive breach in its network\cite{2011_rsa_news} due to malware exploiting a zero-day and an existing Adobe Flash vulnerability to install a backdoor\cite{2011_rsa_hack_how}. Source code of company's SecurID two-factor authentication product was stolen\cite{2011_rsa_damage} with resulting cost of \$66 million for replacing the SecurID tokens\cite{2011_rsa_cost}. 
In April 2011, malware introduced months ago caused a three-day service outage at Nonghyup agricultural bank in South Korea\cite{2011_sk_bank_attack_1,2011_sk_bank_attack_2}. North Korea was blamed for the disruption that prevented 300 million bank customers from using bank ATM's and credit cards\cite{2011_sk_bank_attack_2}.
In April 2011, Sony Playstation Network lost personal information of about 77 million users, including credit card numbers which were stored unencrypted costing Sony \$171 million\cite{2011_sonyps_news,2011_sonyps_exploit,2011_sonyps_damage,2011_sonyps_cost}. During May-June 2011, Sony BMG lost billions of dollars\cite{2011_sonybmg_cost} after it was hacked due to SQL-injection by hacker group LulzSec that posted plaintext data of 50,000 users online to expose weaknesses in the company’s security\cite{2011_sonybmg_news1,2011_sonybmg_news2}
In May 2011, 360,083 credit card account details were stolen in a data breach at Citigroup Inc.\cite{2011_citi_damage,2011_citi_news}. The hack exploited insecure direct object reference, SQL-injection and XSS vulnerabilities\cite{2011_citi_exploit}. Attackers stole \$2.7 million\cite{2011_citi_cost1} and it cost additional \$77 million to the company\cite{2011_citi_cost2}.  
During October 2011, multiple chemical and defense sector companies worldwide came under ``Nitro-attacks'' allegedly carried out by China\cite{2011_nitro_attacks_news,2011_nitro_attacks_report}. Phishing emails with attached malware and remote administration tools were used in these cyberattacks\cite{2011_nitro_attacks_report}. 
In June 2011, International Monetary Fund suffered a cyberattack aimed at stealing confidential information using spear-phishing to install malware. The exact damage caused by the attack remains undisclosed\cite{2011_IMF_motivation,2011_IMF_news}.
In November 2011, a cyberattack employing phishing on Norway's oil, gas and energy systems stole industrial drawings, industrial secrets and user credentials\cite{2011_norway}.

In January 2012, Zappos lost the data of 24 million customers including emails, phone numbers and billing addresses\cite{2012_zappos_news}. 
Reported in June 2012, Gmail accounts of various users were hijacked by unknown state-sponsored attackers that exploited a zero-day vulnerability in Internet Explorer allowing remote code execution\cite{2012_june_gmail_zero_day_IE,2012_june_gmail_zero_day_IE_vulnerability}.
In September 2012, an industrial espionage campaign called \textit{The Mirage Campaign} as it used Mirage remote exploit tool, targeted computers with IP addresses owned by oil, energy and military organizations primarily in Taiwan or the Philippines, with some IPs located in Nigeria, Brazil, Israel, Canada and Egypt\cite{2012_mirage_campaign}. 

In February 2013, water-holing cyberattacks exploiting a zero-day vulnerability in Java targeted Facebook, Apple and Microsoft\cite{2013_feb_facebook,2013_feb_apple,2013_feb_microsoft}. The zero-day exploit was used to automatically download malware. 
In March, a virus from phishing emails caused sudden shutdown of 2 South Korean banks and 3 TV broadcasters, severely affecting broadcasting and ATM services\cite{2013_march_sk}. 
Also in March, \textit{TeamSpy} espionage operation was discovered changing TeamViewer's\footnote{http://www.teamviewer.com/en/index.aspx} DLL files to spy and control targeted computers. The list of victims included high profile industrial, research and diplomatic targets in Hungary and Embassy of NATO/EU state in Russia\cite{2013_march_teamspy}. 
The Reserve Bank of Australia was also hacked in March and malware was installed to gather intelligence on sensitive G20 negotiations\cite{2013_march_rba}. The exact extent of damage remains undisclosed, with China being the alleged attacker\cite{2013_march_rba}.
In April, Japan’s Goo and Yahoo Internet portals were hacked. 100,000 records of user data including financial details like credit card numbers were leaked from Goo.\cite{2013_april_japan}. 
In July 2013, a Ubisoft website was hacked exposing user emails and passwords\cite{2013_july_ubisoft_news}, potentially affecting up to 58 million accounts\cite{2013_july_ubisoft_damage}. JPMorgan Chase bank also suffered a massive data breach in July 2013. 465,000 holders of bank's prepaid cash cards had their personal information accessed by the attackers\cite{2013_dec_jpmorgan}.  
In November 2013, Ireland-based Loyalty Build lost 376,000 credit card numbers and personal information of 1.12 million customers\cite{2013_nov_loyaltybreach_ireland}. 
In December 2013, retail chain Target suffered a massive network breach\cite{2013_dec_target_news}. Attackers installed memory-scraping malware on point-of-sale (POS) devices by gaining entry access to the network using stolen credentials from HVAC service\cite{2013_dec_target_exploit,2013_dec_target_exploit2}. Personal information of up to 70 million people and information on 40 million credit and debit card accounts was compromised. The reported cost of the breach was \$148 million\cite{2013_dec_target_cost}.
In December 2013, Chinese hackers spied on computers of G20 members from Europe before the G20 meeting\cite{2013_dec_g20}. Hackers employed spear-phishing for infecting the targets with malware to gather intelligence on summit negotiations.

\subsubsection{Cyberattacks on Critical Infrastructure} 
\label{subsec:cyberattacks_on_critical_infrastructure}

Critical infrastructure\footnote{http://www.dhs.gov/what-critical-infrastructure} is an attractive target for cyberattacks, given its importance in sustaining normal daily operations. Vulnerabilities in computers supporting critical infrastructure can be equally exploited by cyberattacks as in any other vulnerable system.

In 2003, \textit{Slammer} worm disabled the safety monitoring system at Ohio nuclear plant for 5 hours\cite{slammer_worm_nuke}. In 2004, two Romanian hackers penetrated the network of National Science Foundation's Amundsen-Scott South Pole Station and gained control of the critical life support system, potentially endangering the lives of 58 scientists and contractors\cite{southpole_news_all}. 

In May 2009, FAA's Air-Traffic Network, used to guide and control civilian air traffic is the US, was hacked multiple times because of known vulnerabilities in the system\cite{2009_faa_news1,2009_faa_news2}.

In June 2010, a highly-sophisticated and targeted cyberattack disrupted centrifuges at Iran's Natanz Uranium enrichment plant\cite{2010_stuxnet_damage}. The virus, called \textit{Stuxnet}, exploited 4 zero-day vulnerabilities in Windows operating system\cite{2010_stuxnet_exploit}. Given the sophistication and complexity of targeting a specific system, \textit{Stuxnet} was believed to have been created by Israel and the US to disrupt Iran's nuclear ambitions\cite{2010_stuxnet_attribution1,2010_stuxnet_attribution2,2010_stuxnet_attribution3}. 

In October 2011, a malware with similarities to \textit{Stuxnet} known as \textit{Duqu} was discovered\cite{2011_duqu_symantec,2011_duqu_report}. \textit{Duqu} created back doors which could be exploited to destroy the network at an arbitrary time and also had a keylogger built in to it. A zero-day vulnerability was exploited to distribute \textit{Duqu} trojan\cite{2011_duqu_exploit}. A month later, Iran admitted that its nuclear sites had been hit by \textit{Duqu}\cite{2011_duqu_iran}. In December 2011, cyberattacks on Northwest rail company disrupted railway signals for two days\cite{2012_northeast_rail}. 

In May 2012, \textit{Flame} malware allegedly created by Israel and the US, aimed at slowing down Iran’s ability to develop a nuclear weapon was discovered\cite{2012_flame_news,2012_flame_attribution}. \textit{Flame} exploited existing bugs and a zero-day vulnerability in Windows operating system\cite{2012_flame_attribution,2012_flame_exploit} to infect systems in Iran, Lebanon, Syria, Sudan, the Israeli Occupied Territories and other countries in the Middle East and North Africa two years ago\cite{2012_flame_damage}.
In August 2012, oil producer Saudi Aramco was targeted with \textit{Shamoon} malware to disrupt oil production\cite{2012_saudi_aramco_news,2012_saudi_aramco_motive}. The malware infected 30,000 workstations without disrupting any production. Cutting Sword of Justice claimed responsibility for the attack\cite{2012_saudi_aramco_motive}, though it was attributed to unknown nation-state actor\cite{2012_saudi_aramco_attribution}.

In May 2013, it was revealed that unauthorized access to databases of National Inventory of Dams allowed attackers to get their hands on sensitive information\cite{2013_may_NID,2013_may_NID2}. In the same month, Israel stated that it had prevented cyberattacks from Syrian Electronic Army targeting computers of water systems for city of Haifa\cite{2013_may_israel}.

\subsubsection[Hacktivism]{Hacktivism\footnote{http://dictionary.reference.com/browse/hacktivism}}
\label{subsec:hacktivism}

Cyberspace has also come under attacks motivated by hactivism leading to disruptions and losses in certain cases. Attacks that caused widespread disruptions have only been mentioned here.

In November 2010, hacker group \textit{Anonymous}\cite{2010_operation_payback3} under \textit{Operation Payback} launched targeted DDoS attacks on financial organizations like VISA, MasterCard, PayPal etc. in protest and retaliation to the suspension of WikiLeaks accounts\cite{2010_operation_payback1,2010_operation_payback2}.

In June 2011, \textit{Anonymous} hacked defense contractor Booz Allen Hamilton to publicly humiliate companies and agencies that fail to protect employee and consumer data\cite{2011_BoozAllenHamilton_news}. The attacks were carried out using SQL-injection leaking encrypted passwords and 53,000 .mil email addresses online\cite{2011_BoozAllenHamilton_exploit,2011_BoozAllenHamilton_damage}. 

In February 2013, Bank of America suffered cyberattacks from \textit{Anonymous} with the group claiming that the attacks were in retaliation to bank's online intelligence gathering operation on hacktivists\cite{2013_feb_boa}. Poor security mechanism caused over 6 GB of data to be leaked, including source code for OpenCalais and salary, bonus details of hundred of thousands of executives and employees of various corporations from all around the world\cite{2013_feb_boa}.

\subsubsection{Global Cyber-espionage Campaigns} 
\label{subsec:cyber_espionage_campaigns}


In October 2011, it was reported that 760 organizations worldwide have been under attack by a cyber-espionage campaign stealing sensitive information\cite{2011_oct_760}.

In January 2013, another global malware campaign, called \textit{Red October} was exposed and is believed to have been active since May 2007\cite{2013_redoctober_report}. The campaign exploited vulnerabilities in Java, Microsoft Execl and Word softwares for stealing information from governments, embassies, research institutions, organization in trade and commerce, nuclear/energy research, oil and gas, aerospace and military sectors\cite{2013_redoctober_exploit,2013_redoctober_report}. 

In June 2013, cyber-espionage campaign named \textit{NetTraveler}, allegedly by China, was discovered with victims across multiple sectors including government institutions, embassies, oil and gas industry, research institutes, military contractors and activists in 40 countries\cite{2013_nettraveler_report}.

In September 2013, another cyber-espionage campaign, \textit{Operation IceFrog}, was revealed. It had attacked military, shipbuilding, maritime operations, research companies, telecom operators, satellite operators, mass media and television organizations in South Korea and Japan. The malware exploited known vulnerabilities and hijacked sensitive documents and credentials for accessing internal networks\cite{2013_sep_icefrog}.

\section{Analysis of Cyberattacks} 
\label{sec:trend_analysis}

\begin{figure*}[t]
\centering
\includegraphics[width=1.0\textwidth]{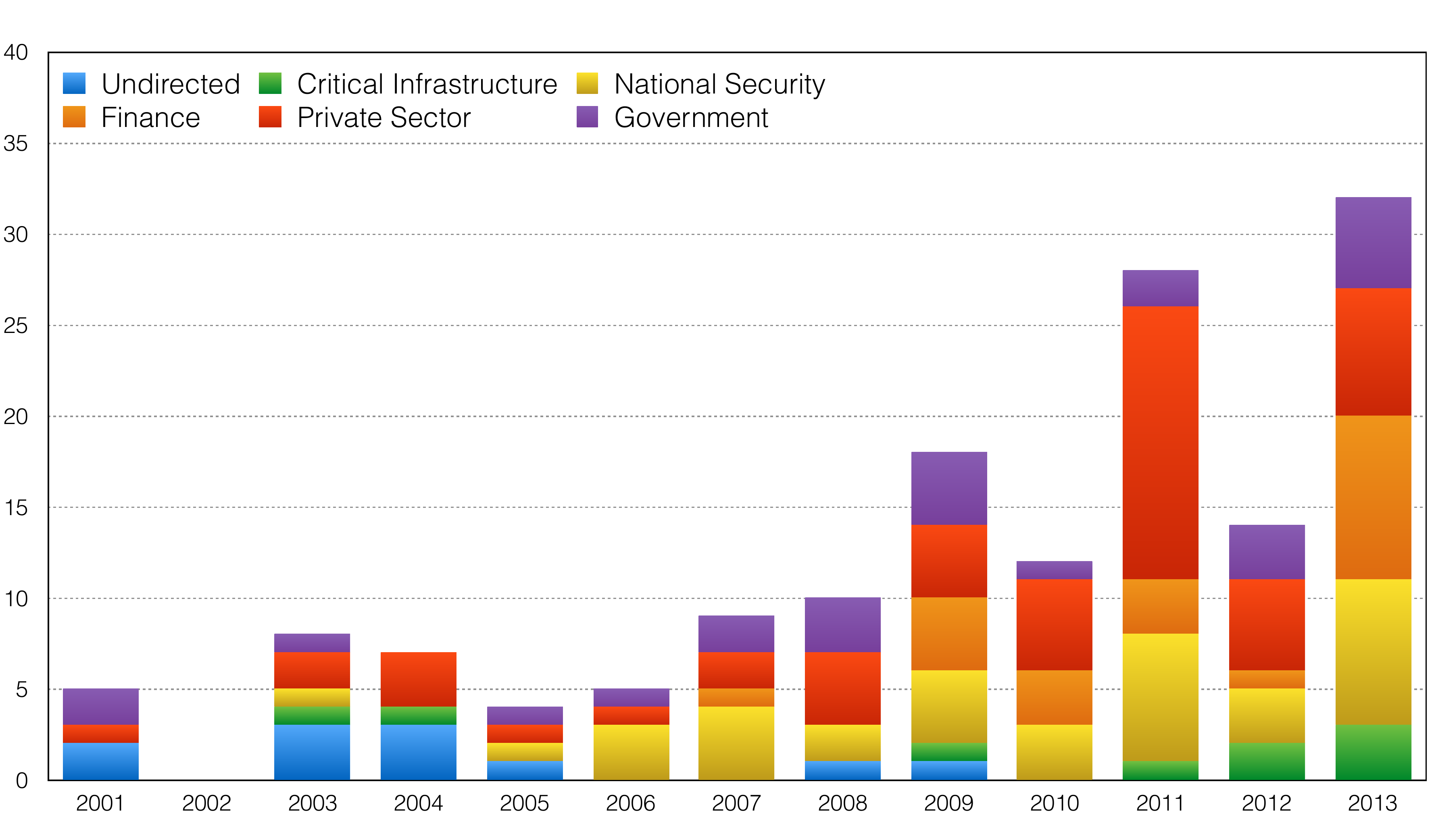}
\caption{Sectors affected by Cyberattacks}
\label{fig:sectors}
\end{figure*}

\begin{figure*}[t]
\centering
\includegraphics[width=1.0\textwidth]{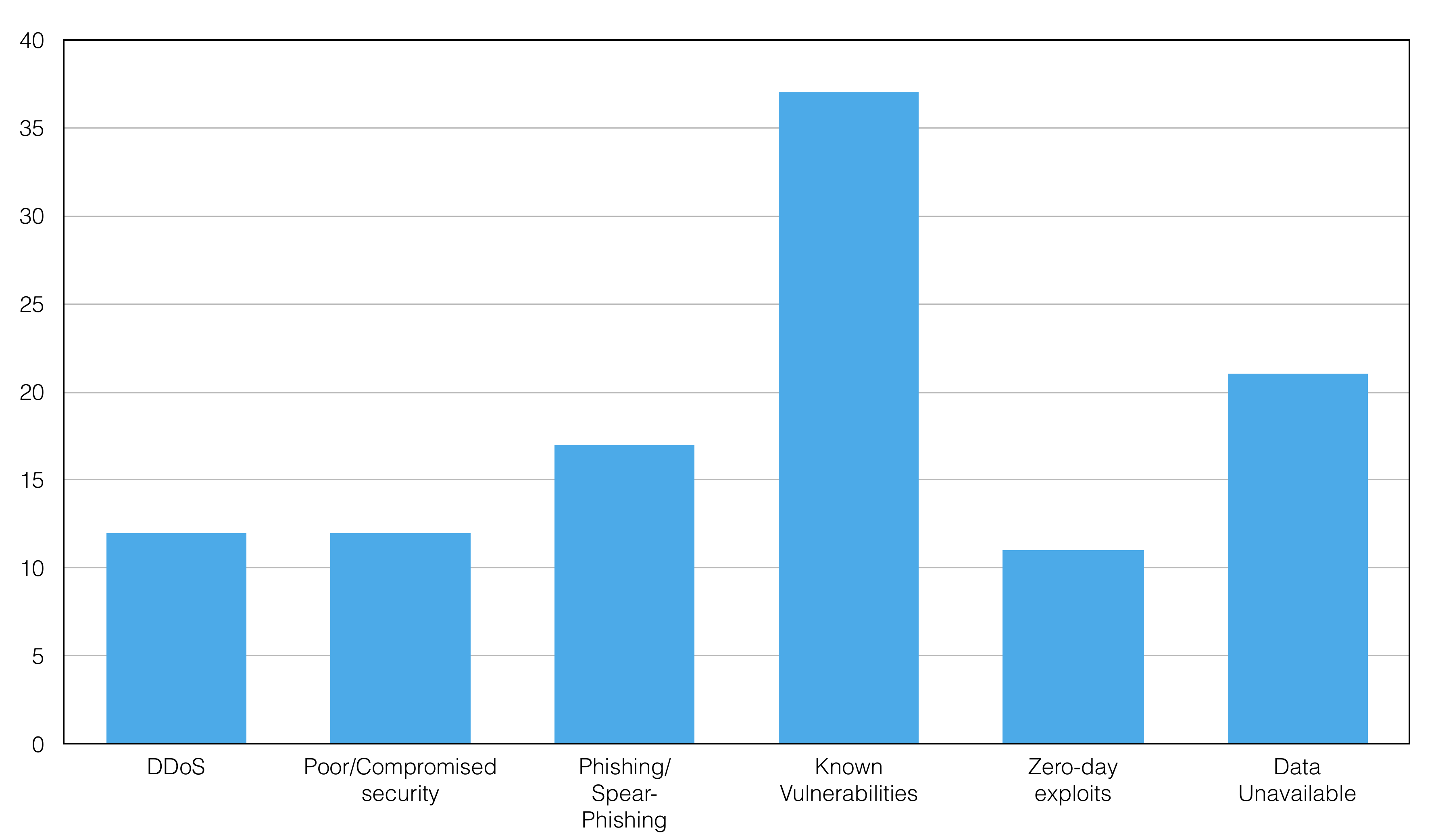}
\caption{Exploits/Techniques employed in Cyberattacks}
\label{fig:exploits}
\end{figure*}

\begin{figure*}[t]
\centering
\includegraphics[width=1.0\textwidth]{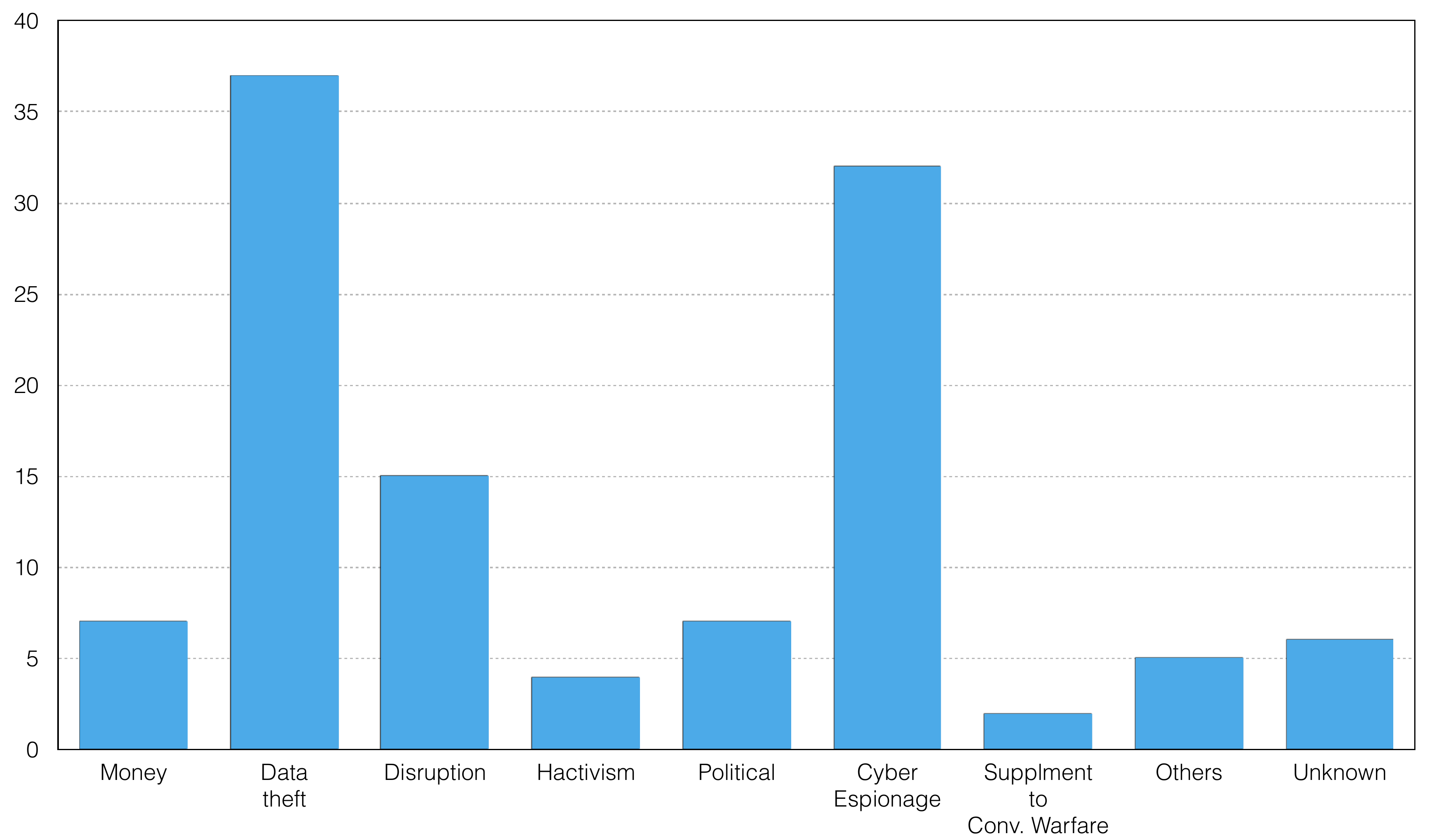}
\caption{Motivation behind Cyberattacks}
\label{fig:motivation}
\end{figure*}

This section provides an analysis of surveyed cyberattacks from various perspectives. Figure \ref{fig:sectors} shows the numbers of attacks and their targeted sectors over the years. The increasing trend in number of cyberattacks can be attributed to adoption of computers in more and more operations and tasks across all sectors. Figure \ref{fig:motivation} sums up the motivation behind various cyberattacks. Techniques and exploits used by the attackers for cyberattacks have been summarized in Figure \ref{fig:exploits}.

Undirected cyberattacks have subsided while targeted attacks have increased and diversified with respect to targeting. Cyberattacks on critical infrastructure\ref{subsec:cyberattacks_on_critical_infrastructure} and national security\ref{subsec:cyberattacks_threatening_national_security} establishments have shown an increasing trend. This should be a major concern for countries who rely on computers and their interconnections for storing sensitive information and proper functioning of critical infrastructure. A continued, sophisticated cyberattack can severely cripple a nation by targeting its critical infrastructure. Private sector companies and organizations have also seen a steady rise in the number of cyberattacks. Governmental organizations as well have fallen prey to well organized and sophisticated adversaries which are going after information, both classified and unclassified, and using the stolen information in future cyberattacks. The attack strategies provide a much more vivid picture to support the argument. 

Figure \ref{fig:exploits} shows that most widespread attacks like Slammer and Blaster worm\cite{slammer_worm_news,blaster_worm_news} and sophisticated global cyber-espionage campaigns mentioned in \ref{subsec:cyber_espionage_campaigns} and many other cyberattacks exploited already known vulnerabilities with patches already available for most of them. Next to known vulnerabilities, poor or compromised security mechanisms paved way for successful cyberattacks. For example, in the cases of Sony\cite{2011_sonyps_exploit} and Target\cite{2013_dec_target_exploit2} breaches, poor security was responsible for the attacks. Compromised security due to attack on RSA\cite{2011_rsa_damage} allowed attacks on Lockheed Martin and Northrop Grumman\cite{2011_lockheed_exploit,2011_northrop_news} in 2011. Phishing/Spear-phishing was the most common mechanism used to deliver malware that exploited vulnerabilities in the system. Information stolen from non-classified sources\cite{titan_rain_news1,titan_rain_times_article,state_dept_news,2008_central_command_real} was most likely used in the phishing attacks mentioned in \ref{subsec:cyberattacks_threatening_national_security}. Distributed denial-of-service attacks remain a popular technique for disruption of services evident from attacks on Estonia\cite{estonia_news2} etc., though the damage caused by them is minimal as compared to other cyberattacks. Zero-day exploits have also surfaced in sophisticated cyber attacks like Stuxnet\cite{2010_stuxnet_damage}. Defending against such attacks is harder, though, measures can be taken be prevent introduction of malware in the system by limiting access and preventing common delivery mechanisms like spear-phishing. Also, exact details of many cyberattacks on companies, governments and other organizations are not disclosed publicly due to security concerns. However, sharing the details can help in developing better security mechanisms and defenses.

Understanding the motivation behind cyberattacks can shed light on the likelihood of a computer system to be targeted by a cyberattack. Data theft shows up as the biggest motivation behind cyberattacks, going after user information, credit card numbers, sensitive information like industrial secrets, corporate access credentials, banking information etc. This indicates that any computer system storing such data is a potential target and therefore, must be secured against all known vulnerabilities and exploits. 
After data-theft, cyber-espionage was the primary motive behind majority of cyberattacks including cyber-espionage campaigns [See \ref{subsec:cyber_espionage_campaigns}] aimed at spying, economic-espionage, industrial-espionage etc. The persistence of these cyberattacks calls for securing all entities in the system to prevent any weak link in the security chain, which also includes the human user.

Disruption of services and networks by DDoS attacks has also motivated many cyberattacks on nation-specific targets [See. \ref{subsec:cyberattacks_on_nations}]. Cyberattacks on Estonia and South Korea\cite{2011_sk_bank_attack_2} should be taken as warnings for future attacks aimed at disruption of services, as they pose threat to stability of daily activities as well as financial losses due to downtime. 
Hactivism can be seen as an emerging threat to computer systems, even though they have been vastly limited to DDoS attacks and defacements, the cost of attacks to the targets can be significant\cite{anon_cost1}.
Criminals have also resorted to cyberattacks, stealing money being one the motives. The number of attacks directly aimed at stealing money remains low as theft of financial information has been covered as data-theft. 
A rising motivation behind cyber-attacks is to supplement conventional warfare. Cyberattacks on Georgia\cite{2008_aug_georgia} and use of cyber-offensive capability used by Israel during a military operation\cite{2007_israel_syrian_hack_news} are among the known incidents to have supplemented conventional warfare with attacks in cyber domain. With sophisticated defense technologies relying heavily on computers systems, protecting those systems against cyberattacks will be paramount in future conflicts.

\section{Conclusion} 
\label{sec:conclusion}

Increasing trend in the number of cyberattacks will continue as more systems get connected to the Internet. Protecting these systems against cyberattacks to ensure normal operation will be the key to minimize disruptions and losses in terms of data, money and time. Majority of cyberattacks we discussed could have been prevented had the systems been kept up-to-date with latest patches. And yet, attacks exploiting known vulnerabilities are not subsiding. This shows that valuable lessons are not being learned from past experiences. Same is the story with attacks employing spear-phishing, which is known to have caused tremendous amount of damage in both classified and unclassified domain. 

Attributing cyberattacks based on technical evidence is also hard due to the very basic structure of the Internet that allows redirection, proxying and spoofing of the source. Alleged sources of most of the cyberattacks described were not based on technical evidence but derived from events in non-cyber world. Therefore, a counter-offensive in response to a cyberattack may not be a feasible option at all and defending the systems in the first place becomes more important.

With increasing adoption of technology and connections of smart devices with the Internet, security of future systems must be considered as an integral part of the system design rather than it being an afterthought. Lack of security in such systems will not only worsen the known consequences but will have far more damaging effects on the society. Learning from past experiences and designing better systems in future can help in changing the trend of increasing cyberattacks.

\subsubsection*{Acknowledgment}

I would like to thank Kyle Ehrlich and Valerie Palermo for reviewing this paper and giving  their valuable feedback and comments.

{\footnotesize
\bibliographystyle{unsrt}
\bibliography{main}
}

\end{document}